\documentstyle[12pt]{article}
\def \F{\phi}

\def \T{\theta}
\def \P{\psi}
\def \D{\delta}

\def \O{{\cal O}}

\def \G{\Gamma}

\def\NP{{\it Nucl. Phys.\ }}

\def\PL{{\it Phys. Lett.\ }}

\def\PRL{{\it Phys. Rev. Lett.\ }}

\def\e{\epsilon}

\def\a{\alpha}
\def\l{\lambda}

\def\s{\sigma}

\def\half{{1\over 2}}

\def\be{\begin{equation}}
\def\eq{\end{equation}}
\def\Tr{{\rm Tr}}
\def\cA{{\cal A}}

\begin{document}

\begin{flushright}
OUTP- 9419P\\
cond-mat/9502118\\
Expanded Jan '95\\
%PACS 36.20.He
\end{flushright}
\vspace{20mm}
\begin{center}
{\LARGE Random Matrix Solution of a Polymer Collapse Model}\\
\vspace{30mm}
{\bf Simon Dalley}\\
\vspace{5mm}
{\em Department of Physics, Theoretical Physics\\
Oxford University, Oxford OX1 3NP, U.K.}\\
\end{center}
\vspace{30mm}
\abstract
A polymer folding model on the square lattice is constructed with attractive
contact interactions of strength $1/c^2$,
$0<c<1$. The corresponding model on a
dynamical random lattice, with freely fluctuating co-ordination number at each
vertex, is formulated as a random two-matrix model and an expression for the
partition function of a length-$L$ chain is derived. Numerical
estimates and analytical
evaluation for  $L\to \infty$  shows a third-order collapse transition at
$c=\sqrt{2} -1$. Geometrical critical exponents are
computed
in each phase and interpreted. The
Knizhnik-Polyakov-Zamolodchikov 2D quantum gravity
scaling relations are used
to predict
the corresponding behaviour on the regular  lattice, which
lies in a different universality class from the
percolation $\Theta$-point of Duplantier and Saleur.

\newpage
\section{Introduction}
\baselineskip .3in
A long polystyrene chain molecule in a solvent
undergoes
a coil--to--globule transition \cite{coil}
 at a finite temperature $T = \Theta$ due to
attractive interactions between different parts of the polymer.
Many theoretical techniques
have been developed to describe this and other processes of
macromolecular folding \cite{dg}, in particular use has been made of
self-interacting
random walks on  lattices such as the
honeycomb lattice \cite{dup} and Sierpinski fractals \cite{dhar}.
In this paper a model  consisting of
folding
chains
 with contact interactions is constructed  on the two dimensional
square lattice.
In order
to study the model it is reformulated on the
ensemble of
two-dimensional simplicial lattices with random fluctuations of the
local
intrinsic curvature, which appear in the study of
two-dimensional quantum gravity \cite{jan,rest}. Statistical
mechanics
on lattices in this ensemble
 represents a
dynamical (annealed rather than quenched) average over a certain class of
fractals. Although
an apparently more complicated problem, the advantage is
that such statistical mechanics problems are often
exactly solvable
using random matrix models \cite{rest}, which yield
expressions for the thermodynamic and some of the geometrical quantities of
interest. The results obtained can be used to infer the
corresponding behaviour on a regular two-dimensional lattice, since
in all of the many
solved examples the qualitative phase
structure of such statistical systems
is the same. Moreover, when the system is conformally invariant on the regular
lattice, there are well-known
techniques
for relating  critical exponents to those of the system on the
fluctuating
lattice \cite{kpz}.

The organisation and main results of this paper are as follows.
In the next section the
polymer model is defined on a square lattice and its relation to
other models which use a different microscopic defintion of contact is
noted. The
dynamical random square lattice is then introduced.
Section 3 explains the random matrix representation
and section 4 describes the standard procedure for solving this
matrix
model using orthogonal polynomials.  In section 5 the properties of a
single polymer are studied, although the construction in sections 3
and 4 is
valid
for a finite polymer fugacity, establishing the exact temperature
$T=T_c$
of a third-order
collapse
transition. By expanding about $T = T_c$ and $T = \infty$, the supposedly
universal exponent $a$ in the asymptotic number of configurations of a
length $L\to \infty$ polymer, $\sim L^a {\rm e}^{bL}$, is determined
to be $a = -5/2$ for $T \to \infty$ and $T \to T_{c}^{+}$
(and presumably for all $T > T_c$) while $a = -3/2$ for $T = T_c$. The
asymptotic behaviour in the $T < T_c$ phase is more subtle but
a value $a=-1$ is obtained for $T \to T_{c}^{-}$.
An attempt to understand the charateristic geometrical behaviour
implied
by these results is made in section 6 by calculation of the dimensions
of the
scaling polymer `star' operators. The phase
structure and scaling dimensions on a fixed regular lattice
are given using the KPZ scaling relation \cite{kpz}. It is concluded
that
$T > T_c$ and $T = T_c$ correspond to the standard dilute and
dense \cite{dense}
phases respectively of a self-avoiding walk in the plane. The case $T
< T_c$ is less clear  but there is  evidence that it entails
an ultra-compact phase whereby the lattice is filled with both polymer and
contact
points.
In any case the collapse transition appears to be in a different
universality
class from the percolating cluster $\Theta$-point of ref.\cite{dup};
the microscopic definition of contact used in this paper seems to
cause
a rather more severe collapse.
Conclusions are summarised in section 7 and a simple co-polymer
model with random sequencing of $n(\P)$ and $n(\F)$ monomers of types
$\P$ and $\F$ respectively, together with
a fugacity for the ratio $n(\F)/n(\P)$ is formulated in an Appendix and
solved with no extra work.

\section{Polymer Model.}

The polymer model is first defined on a fixed square lattice.
A  polymer chain of $L$ steps is a random walk
on the sides (links) of squares which cannot cross itself
(excluded volume), but can occupy a given link any number of times.
Examples of allowed and disallowed configurations are illustrated in figure 1.
Two or more
steps of the polymer ocuppying the same link of the square lattice
incur a contact interaction.
The single polymer partition function is the sum over
all  configurations of the polymer of length $L$ with a weight for
multiple occupation of each occupied link $i$ given by $c^{2-2w(i)}$,
where $w(i)$ is the
number
of steps occupying link $i$ and $0<c\leq1$ is a an attractive
contact  coupling. Note that this
defintion
of contact differs from those often used in studies of self-avoiding
random walks which cannot use a given link more than once.
The one employed here is certainly a less
physical model of the steric repulsion of polymers but leads to an
analytically tractable random matrix theory later.

For the fixed regular square lattice each vertex is
surrounded
by $s=4$  squares. If $s<4$ $(s>4)$ at a vertex
the two-dimensional lattice would
be intrinsically
curved at that vertex, with positive (negative)
Gaussian curvature.
For the dynamical random planar lattice $s$ is allowed to be a
freely fluctuating independent variable at each vertex, subject only
to a fixed total number of $A$ squares in the surface say. In other words
the ensemble of
random fluctuating lattices is obtained  by gluing pairwise
along
links, $A$ squares each of link length $\D$ say, in all possible ways
so as to form an {\em abstract} surface with spherical topology say
(figure 2).
$A$ plays the
role of
and infra-red cutoff on the size of the lattice.
The polymer is a
random
walk on links of squares as before. More generally one can
introduce chemical potentials $\G$ and $g$ conjugate to the number of
polymers $n$ and number of squares $A$ respectively. After multiplying
by $c^{2L}$, the  partition
function
is
\be
{\cal Z} = \sum_{\tilde{w},n,A} c^{2L-\tilde{w}} \G^n g^A
H(A,\tilde{w},n,L) \label{calzed}
\nonumber
\eq
where $\tilde{w} = \sum_{i}( 2w(i)-2)$ and
$H$ is the number of configurations of
$n$ polymers, each of length $L$, with $\tilde{w}/2$ contacts,
on all possible surfaces made from
$A$ squares with spherical topology.
If one imagines $L$ bonds along each side of the polymer which either
connect it to a lattice square or directly to another part of the polymer,
with the normalisation of ${\cal Z}$ chosen in eq.(\ref{calzed})
these have polymer-lattice
$E_{pl}$ or polymer-polymer (contact) $E_{pp}$ bond
energies corresponding to the Boltzmann weights at temperature $T$
\be
{\rm exp}(-E_{pl}/T) = c \ \ , \ \ {\rm exp}(-E_{pp}/T) = 1\label{bolt}
\eq
Thus $E_{pp} =0$ and if $T$ is measured in units of $E_{pl}$, one has
the
correspondence $T = -(\log{c})^{-1}$; hence $c \to 1$ is the high
temperature
and $c \to 0$ the low temperature limit. So far all the geometrical
construction
has been discrete but it turns out that
$H$ grows as $\sim
(1/g_{c})^A$ and therefore there exists a  $g=g_c$ at which
large
area $A$ surfaces become critical and a universal continuum limit can be
established.

\section{Matrix Model.}
In the infinite temperature case $c=1$,
this problem was  mapped onto a random matrix model by Duplantier and
Kostov
\cite{kostov}. In order to analyse  $0< c <1$  a more general
random matrix model will be employed, which is however still solvable.
Let us first recall how the random surfaces described earlier
can be generated from random matrices\footnote{A review of this
technique and  methods of solution can be found in ref.\cite{rev}
for
example}.
Consider the partition function
\be
z = \int \prod_{i=1}^{N}d{\F}_{ii} \prod_{i>j}d{\F}_{ij}d{\F}_{ij}^{*}
{\rm exp}\left( \Tr \left[ -\half \F^2
+{g\over 4N}\F^4 \right] \right) \ .  \label{matrixm}
\eq
where $\F$ is an $N$x$N$ hermtian matrix with elements $\F_{ij}$.
If one expands the integrand in $g$ and performs the
Gaussian
integrals, the expansion can be given a  diagrammatic representation
(figure 2)
of quartic vertices
tied together by the propogators to form a closed graph,
all
possible graphs with $A$ vertices contributing at order $g^A$.
Note
that we are  interested only in connected graphs, so we should take
the
logarithm of $z$.
Because
$\F$ is an $N$x$N$ matrix each graph is also weighted by a power of
$N$. One finds
\be
\log{z} = \sum_{\cal G} g^A {N^{Q-A}\over C({\cal G})}
\eq
where the sum is over all connected closed graphs ${\cal G}$
of co-ordination number
four.
$Q$ is the number of loops in the
graph,
while $C({\cal G})$ is the order of the symmetry group of ${\cal G}$.
These graphs, whose lines are solid in fig.2,
are the dual graphs corresponding to the squares glued along links
described earlier without
any restriction on topology of the closed surface formed. The genus
$G$ of
this  surface is defined by Euler's relation
$v-A = 2- 2G$
where $A$ is the number of square faces
 and $v$ the number of vertices in the random square graph
$\tilde{\cal G}$. Since by duality $Q = v$
and $C(\tilde{\cal G}) = C({\cal G})$, the weight for a genus
$G$ surface is $N^{2-2G}$. As $N\to \infty$, surfaces of
spherical topology $G=0$ dominate and this is the limit which will
always
concern us in this paper.

In order to introduce interacting polymers on the surfaces
consider two random hermitian
$N$x$N$ matrices $\F$ and $\P$ each with the previous measure
(\ref{matrixm})
and
the following partition function
\be
{Z}  = \int {\cal D}\F {\cal D}\P {\rm exp}\left( \Tr \left[ -\half \F^2
- \half \P^2 + c\F \P +\frac{g(1-c^2)^2}{4N}\F^4 +{(1-c^2)^L\over 2} \G
\P^{2L}\right] \right) \ .  \label{matmod}
\eq
The $\F$ matrix generates random surfaces as before, but $\psi$
introduces a new vertex type $\P^{2L}$ with co-ordination number $2L$ and two
new propogators in addition to $\F^2$,
which tie together vertices of the same type (using $\F^2$ or $\P^2$)
or of different types (using $\P\F$) (figure 3(i)).
The propogator weights are
\be
<{1\over N}\Tr \F^2 > = <{1\over N}\Tr \P^2 > = {1\over 1-c^2} \ \ , \
\ <{1\over N}\Tr \P\F > = {c\over 1-c^2} \label{wait}
\eq
The weights have been chosen in order that
in any graph the factors of
$(1-c^2)$ cancel completely between the propogators and vertices, so
$\F^4$ vertices are weighted by $g$ and $\P^{2L}$ by $\Gamma L$.
Forming the dual graphs
once
again, the $\psi^{2L}$ vertices  are interpreted as holes in the
surface
of length $2L$.
The final step in visualizing the polymers is to use the trick of
ref.\cite{kostov} whereby each whole of length $2L$
in the dual graph is sewn up starting at a given point on the edge of
the hole
to form a seam (polymer) of length $L$, illustrated in figure 3(ii).
The extra factor of $L$ in the  weight of a hole
is equivalent to the $L$ possible positions of the given
point which yield a distinct polymer/surface configuration.
Note that the propogator $\P\F$ acts to connect the polymer to the
lattice with weight $c$  while the propgator $\P^2$ acts to
connect one part of the polymer
directly to another part with contact weight one (c.f. (\ref{bolt})).
The $N \to \infty$ limit of $\log{Z}$ eq.(\ref{matmod})
then generates the
required partition function ${\cal Z}$ eq.(\ref{calzed}) on genus zero
surfaces.

\section{Orthogonal Polynomial Solution.}

The matrix model (\ref{matmod}) can be solved in the large $N$ limit
by
the method of orthogonal polynomials \cite{mehta}. Using the
decomposition
of hermitian matrices into unitary and diagonal matrices, the
integration
over the unitary matrices can be explicitly performed
to leave
\be
{ Z}  = \int \prod_{i=1}^{N} d\F_{i} d\P_{i} \Delta (\F) \Delta (\P)
{\rm exp}\left( -
\sum_{i=1}^{N} V(\F_i , \P_i)
\right) \ .  \label{eig}
\eq
where
\be
\Delta (\F) \Delta (\P) = \prod_{i>j}(\F_{i} -\F_j
)(\P_i
-\P_j )
\eq
and
\be
V(\F, \P ) = \half \F^{2} + \half \P^{2} -c\F \P -
\frac{g(1-c^2)^2}{4N}\F^{4} -{(1-c^2)^L\over 2} \G
\P^{2L}  \label{pot}
\eq
with $\F_i$ and $\P_i$  the eigenvalues of $\F$ and $\P$.
Introducing polynomials of degree $i$ and $j$ with first coefficient
normalised to unity: $P_{i}(\F)
=\F^i + \cdots$ , $Q_{j} (\P) = \P^j + \cdots $, where $\F$ and $\P$
are now understood as real variables parameterizing the eigenvalues of the
matrices, they are defined to be  orthogonal under
\be
\int_{-\infty}^{+\infty}
d\F d\P {\rm e}^{-V(\F , \P )}
P_i (\F) Q_{j} (\P) = h_i \D_{ij} \label{ortho}
\eq
Using this property one can show that \cite{mehta}
\be
\log{ Z} =  {\rm const} + \sum_{i=1}^{N-1}
(N-n) \log{f_i} \ , \label{resu}
\eq
where $f_i = h_{i}/h_{i-1}$ and the constant is $g$-independent.
$P$ and $Q$ satisfy recursion relations
\begin{eqnarray}
\F P_{i} (\F) & = & P_{i+1} + r_i P_{i-1} + s_i P_{i-3} + \cdots \\
\P Q_{j} (\P) & = & Q_{j+1} + q_j Q_{j-1} + t_j Q_{j-3}. \label{recur}
\end{eqnarray}
That for $P$ terminates after $L+1$ terms.
One can find a set of equations for $r,s,q,t,$ and $f$ which
can be solved for $f$, and thus $\log{Z}$.
The required equations are
given by the identities
\begin{eqnarray}
\int_{-\infty}^{+\infty}  d\F d\P {\rm e}^{-V(\F,\P)} P_{i-1}' Q_i & = & 0 \\
\int_{-\infty}^{+\infty}
d\F d\P {\rm e}^{-V(\F,\P)} (P_{i}' -i\F^{i-1})Q_{i-1} & =& 0\\
\int_{-\infty}^{+\infty}  d\F d\P {\rm e}^{-V(\F,\P)} P_{i-3}'Q_{i} & =& 0
\end{eqnarray}
and their counterparts obtained by exchanging $P \leftrightarrow Q$
and
$\F \leftrightarrow \P$. By integrating by parts they are explicitly
found to be
\begin{eqnarray}
cq_i & = & f_i [1-{g(1-c^2)^{2}\over N}(r_{i+1} + r_{i}+ r_{i-1})] \\
i & = & r_i -cf_i -{g(1-c^2)^{2}\over N}
[s_i + s_{i+1} + s_{i+2} + r_{i}(r_{i+1}
r_i + r_{i-1})] \\
ct_i & =& -{g(1-c^2)^{2}\over N}f_i f_{i-1} f_{i-2} \\
cr_i & = & f_i - \G \sqrt{f_i} \T [ i,i-1] \\
i & = & q_i -cf_i -\G \sqrt{f_i} \T [i-1,1] \\
cs_i & =& -\G \sqrt{f_{i}f_{i-1}f_{i-2}} \T [i,i-3]
\end{eqnarray}
where
\be
\T [i,j] = L(1-c^2)^L \int d\F d\P {\rm e}^{-V(\F,\P)}
\frac{P_i}{\sqrt{h_i}} \P^{2L-1} \frac{Q_j}{\sqrt{h_j}}
\eq
In the large $N$ limit one can define a continuous variable $x=i/N$
and functions $f_i = Nf(x), r_i = Nr(x), q_i = Nq(x), s_i = N^2 s(x)$,
and $t_i = N^2 t(x) $ so that the previous
equations become
\begin{eqnarray}
cq & = & f -3g(1-c^2)^{2}fr \label{two}\\
x & =& -cf +r -g(1-c^2)^{2}(3s +3r^2) \label{six}\\
ct & = & -g(1-c^2)^{2}f^3 \\
cr & = & f-\G \sqrt{f} \T [i,i-1] \label{one}\\
x & = & -cf +q -\G \sqrt{f} \T [i-1,i]\label{five}\\
cs &  = & -\G f^{3/2} \T [i,i-3] \label{three}\\
\end{eqnarray}
Equations (\ref{three}) and (\ref{six}) will be redundant in this
paper,
while substituting eqs. (\ref{one}) and (\ref{two}) into eq.
(\ref{five})
gives an equation for $f$
\be
x = -cf +{f\over c} -{3g(1-c^2) \over c^2}(f^2 -\G f^{3/2} \theta
[i,i-1]) - \G \sqrt{f} \theta[i-1,i] \ . \label{strieq}
\eq

So far the large-$N$ limit has been taken in order to isolate the
random
surfaces with genus zero, but $g$ is still a free parameter conjugate
to the area $A$. The perturbation series of
eq.(\ref{resu}) in $g$ is actually only convergent for $g<g_c$, which is a
reflection
of the fact that the number of configurations $H$
in
eq.(\ref{calzed}) grows as $(1/g_{c})^{A}$. Therefore by tuning $g \to g_c$
the infinite-$A$ surfaces become critical in the partition function and a
universal continuum limit is attained \cite{rest}
(universal in the sense that one could
equally
have discretised the surfaces with some other type of polygon instead
of squares).
To identify this critical point of the surfaces in the orthogonal
polynomial formalism
it is expedient
to set $\G=0$
(no polymers) and identify the singularity in the solution for $f(1)=f$
of eq.(\ref{strieq}) as $g$ is varied,
\be
1=-cf +{f \over c} -{3g(1-c^2)^{2}f^2 \over c^2}
\eq
yielding
\be
g_c ={1 \over 12} \ , \  f_c = {2c \over 1-c^2}\label{crit}
\eq
As $g \to g_c$ it is useful to define a renormalised (physical) area
of
the surfaces;
if $g=g_c (1-\mu \D^2)$, where $\D \to 0$ is the link length of
squares,
$\mu$
is a renormalised variable conjugate to
renormalised
surface area $\cA = A\D^2$. $\cA$ plays the role of infra-red cutoff
on the area of continuum surfaces for which the ultra-violet cutoff $\D$ has
now been removed. In the continuum limit $\D \to 0$, $\mu$
not only couples to the
universal continuum surfaces ($\cA$ finite) but also to the surfaces finite in
lattice
units $A$ ($\cA \sim O(\D) \to 0$).
The partition function (\ref{resu}) will have a part
regular in
$\mu$, usually representing the latter surfaces, and a non-analytic
part representing the former. The thermodynamic limit is of course the
infinite area one, $\cA \to \infty$.
Now at large $N$ one has from eq.(\ref{resu})
\be
 \log{Z}  = N^2 \int_{0}^{1} dx (1-x) \log{f(x)} + {\rm const.}
\eq
The singular behaviour of $\log{Z}$ arises from $x \to 1$ so it is
appropriate to define scaling variables
\be
x= 1- z\D^2 \ , \ f(x) = f_c ( 1- u(z) \D) \ , \label{scale}
\eq
Then the first non-vanishing order in $\D$ in eq.(\ref{strieq}) occurs at
$O(\D^2)$ giving (at $\G =0$)
\be
u^2 = \mu + z \label{string}
\eq
Hence \cite{rest}
\begin{eqnarray}
 \log{Z}_{\G =0} \equiv Z_0 & = & - N^2 \D^5 \int_{0}^{\infty} dz \ z
u \ + {\rm
regular}\nonumber
\\
& =&  N^2 \D^5 {4\over 15} \mu^{5/2} + {\rm regular} \label{zed}.
\end{eqnarray}
If $u$ is a function of the
combination
$\mu +z$ only, as above, then $N^2\D^5 u(\mu) = - \partial^2 \log{Z} / \partial
\mu^2$ and it has the interpretation of a `susceptibility'.
In the presence of polymers eq.(\ref{string}) is modified by a
$\G$-term; in this paper only the case  of a single polymer
will be addressed in detail.

\section{Single Polymer  Collapse.}

For a single polymer  eq.(\ref{string}) is still appropriate
and one studies the connected expectation value of $\Tr(\P^{2L})$ in this
theory, i.e. $\log{Z}$ expanded to first order in $\G$, $\log{Z}
= Z_0 + \G Z_1 +
\cdots$. In terms of orthonormal polynomials, connected one-point
Green's functions of the matrix model are given by
\begin{eqnarray}
Z_1  & = & Z_0 < (1-c^2)^L \Tr ( \P^{2L}) >_{\G=0} \nonumber \\
& =& (1-c^2)^L \sum_{i=1}^{N} \int d\F d\P {\rm e}^{-V(\G=0)}
{P_i \over \sqrt{h_i}} \P^{2L} {Q_i \over \sqrt{h_i}} \label{onorm} \ .
\end{eqnarray}
In order to evaluate the last quantity it is illuminating to introduce
 raising and lowering operators on the orthonormal polynomials defined by
\be
a^{\dagger} \left[ {Q_i \over \sqrt{h_i}}\right] = \left[ {Q_{i+1} \over
\sqrt{h_{i+1}}} \right] \ \
, \ \ a\left[ {Q_i \over \sqrt{h_i}} \right] = \left[ {Q_{i-1}
\over \sqrt{h_{i-1}}} \right] \ .
\eq
Then from (\ref{recur})
\begin{eqnarray}
\P \equiv \sqrt{f} a^{\dagger} + {q \over \sqrt{f}} a + {t \over
f^{3/2}}a^3 \label{psi}
\\
cq = f\left(1-{3g(1-c^2)^2 f \over c}\right) \ , \ ct = -g(1-c^2)^2 f^3
\end{eqnarray}
Thus by orthogonality, $\P^{2L}$ contributes in (\ref{onorm}) only when there
are
equal numbers of $a$ and $a^{\dagger}$ in its expansion. Defining
$Z_1 = \sum_{i=1}^{N} Z_{1}'$,
one finds (as always, for $N \to \infty$)
\begin{eqnarray}
Z_{1}'(L,c,f,g) & = & (1-c^2)^L (2L)! f^L
\sum_{p=0}^{[L/2]} {[ c -3g(1-c^2)^2 f]^{L-2p}
[-g(1-c^2)^2 f]^{p} c^{3p -2L} \over (L-2p)! (p+L)! p!}
\label{sum}\\
{(1-c^2)^2 gf \over c} & =& {1-c^2 \over 6} \left[1- u\D +
O(\D^2)\right]
\nonumber
\end{eqnarray}
where $u$ is given by eq.(\ref{string}) and $[L/2]$ indicates
greatest integer $\leq L/2$. This formula is the first main result.

If, as the continuum limit $\D \to 0$ is taken, $L$ remains finite,
the
polymer will appear merely as an infinitesimal puncture of length $2L$
in the
surfaces
of finite renormalised area ${\cal A}$. In this case eq.(\ref{sum})
reduces to the form
\be
Z_1 = N \int_{0}^{\infty} dz\  \left[ X(L,c) - Y(L,c)u\D +
O(\D^2)\right]
\label{dot}
\eq
The parts regular in $\mu$, including the first term, represent the
non-universal contributions due to the polymer on finite-$A$
surfaces. The
non-analytic part at lowest order in $\D$ gives the scaling
behaviour of the puncture operator one-point function on the finite-${\cal A}$
surfaces in the continuum limit
\be
Z_1 = {\rm regular} + {2 \over 3} N \D^3 Y(L,c) \mu^{3/2}
\eq
This is basically $\partial Z_0 / \partial \mu$ (c.f. eq.(\ref{zed}))
since a puncture can be anywhere on the surface and therefore measures
its
area $\cA$ conjugate to $\mu$. Note that the overall power of $N$
occurs
because Euler's relation introduced earlier is more generally $v-A
=2-2G - h$
where $h$ is the number of holes in the surface.
Changing $L$ or $c$ does not change the scaling dimensions and there
are
no singularities of $Y(L,c)$,  so no phase
transitions.

In order that the polymer remain  an extended object on the surfaces of
finite
$\cA$ one must scale its length as $L = l \D^{-2/\nu D}$, for some
appropriate
exponent $\nu D$. $D$ is understood as some intrinsic fractal
dimension of the lattice which would be two if the lattice were
regular,
while $\nu$ is the usual mean square size exponent of a polymer.
In order to find $\nu D$, and in particular its dependence
upon $c$, one must evaluate the large $L$ behaviour of the sum
(\ref{sum}).
This is
not
straightforward because the terms in the sum alternate in sign. For the case
$c=1$ \cite{kostov},
which
corresponds to infinite temperature, only the
first
term survives, leading to the
result $\nu D =1$  and
\be
Z_1' = {2^L (2L)! \over (L!)^2} {\rm exp}(-ul) + {\rm regular}
\eq
in the continuum limit $\D \to 0$.
The large $L$ behaviour of the prefactor is $\sim 2^L /\sqrt{L}$,
showing the characteristic (non-universal) exponential growth of the number of
configurations,
and that for $Z_1$ the  part non-analytic in $\mu$
grows as $\sim
2^L / L^{5/2}$. The exponent $5/2$ is expected to be universal
and one of
the main objectives is to find it's dependance upon $c$. Implicitly we
are therefore
assuming that this characteristic form of the large $L$ behaviour
persists as $c$ is reduced from one, i.e. when $L = l/\D$
\be
Z_{1}' = G(L,c)[ {\rm e}^{-\l (c) ul + O(\D)} + O({\rm
const}^{-O(\D^{-1})})] \label{asym}
\eq
where
$G \propto L^a {\rm e}^{b(c) L}$ and $a$ is universal. As always, by
universal
it is meant independent of the form of discretisation of the surface,
rather
than independent of the microscopic definition of contact interactions.
The author has
not managed to {\em derive} the asymptotic form (\ref{asym}) except
in the neighborhoods of $c=1$ and $c=\sqrt{2}-1$, which
will
be discussed shortly.

It is  however useful for orientation to first investigate the exact formula
(\ref{sum})
numerically at large but finite $L$ and fit it to the form (\ref{asym}).
Evaluation of (\ref{sum}) for increasing $L$ indeed
confirms exponential growth with $L$ for general $c$.
Figure 4  plots $\l (c)$ obtained from
\be
ul\l = \log{\left[ {Z_{1}'(L,c,0) \over Z_{1}'(L,c,ul)} \right]}
\label{derive}
\eq
where $Z_1'$ is evaluated for $L=150$ --- corrections to the right hand side
above are $O(\D) \sim O(1/L)$.
$b(c)$ does not depend upon $\mu$ and so represents some short
distance
lattice artifacts.
On the other hand $u\l$ is expected to be universal and
represents an induced free energy per unit
length of the  polymer due to fluctuations of the
finite-$\cA$
surfaces;\footnote{Note that it is the presence of an extra
dimensionful
parameter $\mu$ on the fluctuating lattice which enables one to
extract a universal part from the free energy of the polymer,
which in the case of
a dilute polymer is a like a boundary free energy.}
it is non-analytic in $\mu$ from eq.(\ref{string}).
The
behaviour of $\l (c)$ thus governs the interesting geometrical
properties
of the polymer. In particular, fig.4 seems to show a transition
to $\l (c) =0$ at $c \sim 0.4$, which will be verified shortly.
Physically,
 as $c$ is reduced the polymer favours polymer-polymer rather than
polymer-lattice bonds and  the induced free energy $\l$ falls.
Below a certain $c_c \approx 0.4$ the polymer collapses entirely and
is
only connected into the surface by a small number of polymer-lattice
bonds. This is the coil-to-globule transition on fluctuating random
surfaces. The polymer-polymer contacts also induce free energy per
unit
length but at a lower order in $\D$, and it will be necessary to
choose
a different scaling exponent $\nu D$ in order to see this contribution
near $c = c_c$.

In order to find $c_c$ exactly, determine the  exponents $a$ and $\nu D$,
and
the details of the purported phase transition, it is useful to
make the replacements $a \to {\rm e}^{ip}, a^{\dagger} \to {\rm
e}^{-ip}$ in $\P$ (\ref{psi}),
enforcing the cancellation of $a$'s and $a^{\dagger}$'s by integrating
over
$p$ thus (c.f. eqs.(\ref{onorm})(\ref{psi}))
\be
Z_{1}' = (1-c^2)^L \int_{-\pi}^{\pi} {dp \over 2\pi} \left[ \sqrt{f}
{\rm e}^{-ip} + {q \over \sqrt{f}}{\rm e}^{ip} + {t\over f^{3/2}}{\rm
e}^{3ip}
\right]^{2L} \label{pees}
\eq
As the continuum limit  $\D \to 0$, $L \to \infty$, is approached one
would like
the integrand to exponentiate simply using the property $(1-\beta/L
)^{2L} \to  {\rm e}^{-2\beta}$ ($\beta$ finite). To see why this does
not
happen in general it is expedient to
expand $\P$  in $\D$ and $p$, using previous definitions for $f,q,$ and
$t$ in terms of scaling variables $u$ and $\mu$;
\begin{eqnarray}
\P  & \equiv & {1\over \sqrt{(1-c^2)2c}}[a_1 +a_2 ip+ a_3 u +a_4 p^2
+ a_5 iup \D + a_6 u^2 \D^2 +a_7 \mu \D^2 \nonumber
\\ &&
+ a_8 ip^3  + a_9 u p^2 \D
+ a_{10} iup^3\D
+ a_{11} p^4 + \cdots]
\label{long}
\end{eqnarray}
The coefficients $a_i(c)$ are easily computed, in particular
\begin{eqnarray}
a_1 & = & {1\over 3}(3 +5c + 3c^2 +c^3) \\
a_2 & = & 1 -3c + c^2 + c^3   \\
a_3 & = & \half (1-c-3c^2 - c^3)     \\
...& & \ldots {\rm etc.}\nonumber
\end{eqnarray}
This and subsequent algebraic computations were done by hand and
checked
by computer.
Since $a_2 \neq 0$ in general, this gives the $\beta$ above an (infinite)
imaginary part and the resultant oscillatory behaviour means that one
must keep all orders of the expansion in $p$ (\ref{long}) in order to
even begin to evaluate the integral (\ref{pees}). This problem will
turn
out to be  obviated if
$a_2 = 0$ for some
value
of
$c$ since one can then easily separate off the contribution to the
$p$-integral non-analytic in $\mu$, coming from the $p \sim
O(\D^{1/\nu D})$
region.
In the interval $0 \leq c \leq 1$,
$a_2$ has roots at $c=1$ and $c=\sqrt{2} -1$, the latter shortly being
interpreted as the collapse temperature $c_c$. The strategy now is to expand
$c$ about these two roots.

First consider the neighborhood of $c=1$ (high temperature limit).
In order to reproduce the $c=1$ solution \cite{kostov} one finds that
for $L = l/\D$ ($\nu D =2$) the
contribution of the integral (\ref{pees}) non-analytic
in $\mu$
comes from the region  $p \sim O(\sqrt{\D})$.
Writing $c=1-\e$ for small $\e$, $a_2 = O(\e)$
and working perturbatively to lowest
order
in $\e$ yields
\begin{eqnarray}
\P^{2L} & \equiv & ((1-c^2)2c)^{L} \left[a_1 +
\left(a_3 +{a_2 a_5 \over 2
a_4} \right) u  \D + \left( a_4 + {3 a_2 a_8 \over 2 a_4}\right)
\tilde{p}^{2} \D
+  \cdots\right]^{2L} \label{lung} \\
\sqrt{\D} \tilde{p} & = & p + i\sqrt{\D} \rho \ \ , \ \
\rho = {a_2 \over 2 a_4
\sqrt{\D}}
+ {a_5 u \sqrt{\D} \over 2 a_4}+ O(\e^2,\e u\sqrt{\D}) \nonumber
\end{eqnarray}
where ellipses indicate either  $O(\D^{3/2})$ terms, which do not
survive
the continuum limit, or terms which renormalise the displayed
coefficients
in (\ref{lung}) at $O(\e^2)$.
Therefore
\begin{eqnarray}
Z_1'  & = & {a_{1}^{2L} \sqrt{\D} \over 2\pi (2c)^{L}}
\int_{i\rho - \pi/\sqrt{\D}}^{i\rho + \pi/\sqrt{\D}}
d\tilde{p} \ {\rm exp}\left[ 2l \left({a_3\over a_1} +{a_2 a_5 \over
2 a_1 a_4} \right) u  + 2l \left( {a_4\over a_1} + {3 a_2 a_8 \over
2 a_1
a_4}\right) \tilde{p}^{2}
+ \cdots \right] \\
& = & \sqrt{{\D}\over 4\pi l} [8 + 0(\e)]^L  {\rm
exp}\left[-ul\left(
1-{4\e \over 3} + O(\e^2)\right) \right] + {\rm regular}\label{oh}
\end{eqnarray}
in the continuum limit $\D \to 0$, where the integral
\be
\lim_{\D \to 0} \int_{i\rho - \pi/\sqrt{\D}}^{i\rho + \pi/\sqrt{\D}}
dx \
{\rm e}^{-x^2}
= \sqrt{\pi} \ ; \  \lim_{\D \to 0} \sqrt{\D} \rho < \pi
\eq
has been used, valid since $\rho \sqrt{\D} \sim O(\e) << 1$ by hypothesis.
The same result can also be derived  by
expanding
the formula (\ref{sum}) in $\e$.
Eq.(\ref{oh}) is of the form (\ref{asym}) with a $\l$
that
correctly matches onto the $L=150$ curve of fig.4  near $c=1$.
The partition
function
itself would then be
\begin{eqnarray}
Z_1 & = & N\D^2 \int_{0}^{\infty} dz \ Z_1' \label{ah} \\
   &\propto & N\D^{5/2}   {{\rm const}^L
\over \l l^{3/2}} \left( \sqrt{\mu} + {1\over \l l}\right)
{\rm e}^{-\l l \sqrt{\mu}} + {\rm regular} \label{form}
\end{eqnarray}
using eq.(\ref{string}). Large area $\cA \to \infty$ corresponds to
$\mu \to 0$, in which case the  power exponent is $a=-5/2$.

Fortunately there is one other value of $c$ where $a_2$
 vanishes, given by $c=\sqrt{2} -1$; $a_3$ also
vanishes at this point. Writing $c = \sqrt{2} -1 +\e$ for small $\e$
\begin{eqnarray}
a_2  & = & 4(1-\sqrt{2})\e + (3\sqrt{2} -2)\e^2 + \e^3 + O(\e^4) \nonumber \\
a_3 & = & -2\e -  {3\sqrt{2} \e^2 \over 2} -{\e^3\over 2} + O(\e^4) \ .
\nonumber
\end{eqnarray}
All other $a_i$'s in (\ref{long}) are non-vanishing at $c= \sqrt{2}-1$
and
it is straightforward to evaluate their $\e$-expansions and substitute
in (\ref{long}) once again. The result to lowest non-zero
order in $\e$, which
requires
one to consider all terms displayed in eq.(\ref{long}), is after a
tedious
calculation
\be
\P^{2L} = \left({a_1\over \sqrt{(1-c^2)2c}}\right)^{2L}
{\rm exp} [ -\a \e^3 ul - (2 a_4/a_1) \tilde{p}^2 l
+ \cdots ]
\label{ohlong}
\eq
where
ellipses are $O(\D^{1/2})$ terms
which do not contribute to the continuum limit or terms which
renormalise
the displayed coefficients beyond the leading non-zero order in $\e$.
$2 a_4 /a_1 = 9 + O(\e)$ while $\a$ to lowest order in $\e$
is given by
\begin{eqnarray}
\a \e^3 & = & \e \left({a_2 a_5 \over a_4 a_1} +2{a_3 \over a_1} \right)
- \e^2 \left( {3a_8 a_{2}^{2} a_5 \over 4 a_{4}^{3} a_1} + {a_9
a_{2}^{2}
\over 2 a_{4}^{2} a_1} \right) \nonumber \\
&&- \e^3 \left( {a_{10} a_{2}^{3} \over
4 a_{4}^{3} a_1} - {a_{11}a_5 a_{2}^{3} a_5 \over 2 a_{4}^{4} a_{1}}
+ {3 a_8 a_{2}^{2} \over 8 a_{4}^{5} a_1} ( 3a_8 a_2 a_5 + 2 a_4 a_2
a_9 )\right) + O(\e^4)
\end{eqnarray}
Numerically $\lim_{\e \to 0} \a \approx 35$.
Once again the appropriate
scaling behaviour was $L = l/\D$.
Performing the
$\tilde{p}$-integral and using (\ref{string}) the partition
function is
then
\be
Z_1 \propto {\rm const.}^L N\D^{5/2} {1\over \sqrt{l}}
\int_{\mu}^{\infty}dy\ {\rm e}^{  -\a \e^3 \sqrt{y} l} + {\rm regular}
\label{comp}
\eq
($u = \sqrt{\mu + z} \equiv \sqrt{y}$) which is again of the form
(\ref{oh})
with
\be
\l  = \a \e^3 + O(\e^4). \label{vanish}
\eq
The calculation for large but finite $L$ shown in fig.4 approaches
this limiting behaviour with increasing $L$. At $\e =0$ there  is a
collapse transition because as $\l \to 0$
\be
Z_1 \to {\rm regular} + {1\over 3} N\D^{5/2} {\rm const}^L \l
\sqrt{l}
\mu^{3/2}
\eq
which is of the same form as the puncture operator one-point function
(\ref{dot}) -- the polymer has collapsed to a point.
Since $\l \sqrt{\mu}$
represents the (non-analytic part of the)
free energy per unit length of the polymer due to
its connections into the lattice, its vanishing (\ref{vanish}) can be
interpreted as a third-order collapse transition if $\l$ is
identically zero for $c< \sqrt{2} -1$. (This is consistent with the
evaluation of the exact result (\ref{sum}) for $L=150$ in fig.4.)
For negative $\e$ the integral for the partition function (\ref{comp})
is not convergent.
This is because the higher orders in $\D$ have been
neglected, i.e. there are corrections to the argument of the exponential
integrand $ly\D, ly^{3/2}\D^2, \ldots$
which become
important at $y \sim
O(1/\D^2)$, contributing to the regular part.
In terms of the discrete variable used earlier,
$i = Nx$, the sum
$Z_1 = \sum_{i=1}^{N} Z_1'$  no longer has a singular part in $\mu$
coming
from the  finite-$i$
region but only the regular part coming from $i \sim O(N)$.
In fact the result of the $y$-integral becomes, to leading order in
$\D$, some $\mu$-independent constant since it is insensitive to the
lower
limit and $\mu$ enters explicitly in the argument of the
exponential only at lower order in $\D$ (terms like $l\mu \D,
l\mu\sqrt{y}\D^2, \ldots$).
Therefore $\l = 0$ for $c = \sqrt{2} -1 + \e$, perturbatively in $\e <0$.

In order to make the polymer an extended object once again when $c
\leq \sqrt{2} -1$ a different exponent $\nu D$ must be
chosen. Consider
first  the
collapse point itself, $c= \sqrt{2} -1$. The order $u$ term in $\P$
(\ref{long})
vanishes and one must consider the next order in $\D$, that is $\mu$
and $u^2$ terms, which gives
\be
\P^{2L} \equiv \left({a_1\over \sqrt{(1-c^2)2c}}\right)^{2L}
{\rm exp} [(\mu -3u^2/2)l  -  9 \tilde{p}^2 l  + \cdots ] \ .
\label{olong}
\eq
Here
$L = l/\D^2$ ($\nu D =1$) has been chosen to pick up the lower order
terms $\mu$ and $u^2$ in the continuum limit and $p = \D
(\tilde{p} - i\rho / \sqrt{\D})$ has been rescaled accordingly;
ellipses are higher orders in $\D$.
Then
\begin{eqnarray}
Z_1 & \propto &{\rm const.}^L N\D^3
\int_{\mu}^{\infty}dy \int_{i\rho/\sqrt{\D} - \pi/\D}^{i\rho/\sqrt{\D}
+ \pi/\D} d\tilde{p}  {\rm e}^{
-9 \tilde{p}^{2} l -3 yl/2 + \mu l} + {\rm regular} \nonumber \\
 & = & {2 \over 3} {\rm const.}^L N\D^3 {{\rm e}^{ -\mu l/2 }\over {l}^{3/2}}
+ {\rm regular} \label{compb}
\end{eqnarray}
The finite-$i$ region now also gives something analytic in $\mu$, but
in
this case nevertheless represents
{\em finite}-$\cA$ surfaces. The $\cA = 0$ surfaces
--- $\cA \sim O(\D^2)$ more precisely --- are characterised by the
fact
that a finite number of derivatives with respect to $\mu$ can remove
their
contribution, since each derivative brings down a factor of $\cA$.
The first term in (\ref{compb})
cannot be removed by differentiating. It can be
interpreted
as due to  surfaces dense with polymer whose area is proportional to the
polymer
length, which is now conjugate to $\mu$;
for small $l$ differentiating does remove its
contribution.
The result $a = - 3/2$ and $\nu D = 1$ agrees with a
dense
phase of polymers constructed in ref.\cite{kostov}.
This is discussed futher in the next section.

For $c < \sqrt{2} -1$ the author has not been able arrive at a
wholely satifactory
quantitative understanding of a scaling limit with fixed polymer
length
$L$. It was shown
above
that $Z_1$  no longer has an obvious universal contribution from
the finite-$i$ region. The precursor to this was already seen at $c
=\sqrt{2}
-1$ where the finite-$i$ region gave something regular in $\mu$ but
nonetheless had an interpretation in terms of scaling surfaces and
polymer. The origin of the difficulty seems to lie in the fact of
working
with a polymer of fixed length $L$ which is dense on the surfaxe.
To resolve  ambiguity and identify a
part
singular in $\mu$ it turns out to be necessary to sum over $L$ with
respect
to a monomer fugacity $K$. This will be done in the next section in
the more
general context of scaling operator two-point functions.

One notes however that the polymer
in the low temperature phase should  to be in an
ultra-compact
state due to the allowance of multiple occupation of links in the
model.
This is highlighted in the low temperature expansion;
expanding (\ref{sum}) about $c=0$, only a finite number of terms survive
to
given order of $c$.
The result is
\be
Z_1' = {(2L)!\over (L!)^2} \left[ 1 + {c^2L\over 3}(1-2u\D +
O(\D^2,1/L)) + O(c^4 L^2)
%+ {5c^4L^2
%\over 18}(1 - {36 u \D \over 5} + O(\D^2,1/L)) \cdots
\right]
\eq
These terms  have a clear geometrical meaning. At $O(c^0)$
the  term $(2L!)/(L!)^2 \sim 2^{2L}/\sqrt{L}$ corresponds to the
number of ways of folding the polymer so that it has no connections
into the surface at all. Each subsequent  power of $c^2 L$ corresponds
to breaking  a polymer-polymer bond, of which there are
$L$,
and inserting the dangling bonds into the dual graph on a $\F^2$
propogator.
For example,  order $c^2$ corresponds to folding the polymer entirely
onto {\em one} of the links of the square lattice; equivalent
in the dual graph to insertion of a $\Tr
\F^2$
puncture operator times a combinatorial factor to account for the
number of ways of pairing the $2L-2$ other bonds of the polymer.
Indeed,
${\partial \over \partial \mu} < \Tr \F^2 > = 1- 2u\D + O(\D^2)$,
where
the leading non-analytic term corresponds to two punctures on the
universal
finite $\cA$ surfaces, one from the polymer and the other from
$\partial
/\partial \mu$ (compare with eq.(\ref{dot})).

\section{Scaling Operators.}

In order to better appreciate the geometrical implications of the
calculations
just performed and convert the results to the corresponding ones on a
fixed regular lattice it is expedient to derive the scaling dimensions
of
the usual polymer `star' operators. These are the operators $\{
\O_{S}
\}$  which act
as
sources for $S$ polymer lines; the two-point functions $<\O_{S} \O_{S} >$
are watermelon networks illustrated in figure 5. These configurations
are
most simply discussed by introducing a variable $K$ conjugate to the
total
length $L$ of all polymers in the configuration, and in the case of a
fluctuating lattice the operators should be integrated over the
lattice
because there is no translational invariance. Thus one is considering
the
correlators
\be
< \int \O_{S} \int \O_{S} > = {1\over Z^0} \sum_{L=1}^{\infty} K^L
Z^{S}
(L) \label{star}
\eq
where $Z^{S}(L)$ is the partition function for the $S$-line watermelon
network with fixed total length $L$ of all the polymers. In particular
for $S = 1$ one has $Z^1 = Z_1$ calculated in the previous
section. In a dilute phase of polymers
one also expects $Z^0 = Z_0$ (\ref{zed}) since the
polymers
have negligible backreaction on the background lattice. In a
dense ($\nu D \leq 1$)
phase however it is incorrect to normalise correlation functions with
$Z_0$
since the dense polymer alters significantly
the background geometry. In general
one should normalise by
a  fluctuating lattice partition function with some
scaling exponent $Z^0 \sim (\D^2)^{2-\gamma_{str}}$ which takes into
account any backreaction. Fortunately it can be determined
by an independent argument \cite{dup}.
The `gravitational' scaling dimension $\Delta_S$ of $\O_S$
is given by the dependence on the lattice spacing
\be
< \int \O_S \int \O_S > \sim (\D^2)^{2\Delta_S -2} \ .
\eq
$\Delta > 2$ and $\Delta <2$ imply irrelevance and relevance in the
continuum limit respectively.
These dimensions are related to the conformal weights $\Delta^{(0)}$
of operators in
a
conformal field theory in the plane \cite{bpz}
of central charge $c$ by the KPZ
formula \cite{kpz}
\begin{eqnarray}
 \Delta^{(0)}& =& \Delta \left(1 - {1- \Delta  \over \kappa} \right)
\label{ddk}\\
c  & = & 1-6{(1-\kappa)^2 \over \kappa} \ , \ \kappa > 1
\end{eqnarray}
In general the exponent $\gamma_{str}$ does not determine $c$ uniquely unless
some
further information is available. For the problem studied in this
paper the charateristic behaviour of $\Delta^{(0)}$, that is whether
it
is negative, zero, relevant, marginal, or irrelevant, follows that
of the corresponding $\Delta$.

The exponent $\Delta_1$ is related to the
power exponent in the single polymer partition function $Z_1 \sim {\em
e}^{bL} L^a$
\be
2\Delta_1  =  \gamma_{str} - {a+1\over \nu D}
\eq
Similarly $ \Delta_{1}^{(0)}$ is related to the corresponding power
exponent in the growth of the single polymer partition function in the
plane, conventionally written $L^{\gamma -1}\mu^L$, by
$\gamma / \nu = 2-2\Delta_{1}^{(0)}$
where the latter $\nu$ is the
usual mean-square-size exponent $R^2 \sim L^{2\nu}$. Since $\O_2$
represents marking a point on the polymer, it couples to the length
$L$
and hence there exists are relation to $\nu D$ given by \cite{kostov}
\be
\nu D = {1 \over 1 - \Delta_2} \label{new}
\eq
with a similar relation in the plane on setting $D=2$.
The higher even operators $\O_{2n}$ for $n >1 $ represent a contact
between $n$ different parts of the polymer simultaneously.

To illustrate the calculation of the star operator two-point functions
(\ref{star}) consider the case $S=2$. The relevant configurations are
those
of a closed loop of polymer with two marked points on the
polymer. This
constructed from two pieces of surface $S_1$ $S_2$ (figure 6), each
with disc topology and the same boundary length $L$. The polymer is
formed
by first pinning together the two boundaries at (arbitrary) given
points
$P_1$ and $P_2$ on the boundary of $S_1$ and $S_2$ respectively, then
sewing the disc boundaries together. Finally one chooses another
(arbitrary)
given point $P_3$ on the seam. The total amplitude for these
configurations
is $Z^2(L) = L Z_1(L/2) Z_1(L/2)$, that is two independent disc
amplitudes, one  with two marked points the other with one marked
point on the boundary.
Using the integral representations (\ref{pees})(\ref{ah}) one
therefore has
\be
Z^0 < \int  \O_2 \int \O_2 > = {N^2 \D^4 \over (2\pi)^2}
\int_{0}^{\infty}
dz_1 dz_2 \int_{-\pi}^{\pi}dp_1 dp_2 {1 \over (1-K\P_1 \P_2)^2}
\eq
Generalising the construction to sewing $S$ discs together to form the
watermelon network (fig.5) gives
\begin{eqnarray}
Z^0 < \int  \O_S \int \O_S > & = &\left({N \D^2 \over 2\pi}\right)^S
\int_{-\pi}^{\pi}dp_1\ldots dp_S \int_{0}^{\infty}
dz_1\ldots
dz_S {1 \over (1-K\P_1 \P_2)}{1 \over (1-K\P_2 \P_3)}\ldots \nonumber \\
&& \ldots {1 \over (1-K\P_S \P_1)} \label{conv}
\end{eqnarray}
where hereafter the rescaling $\sqrt{1-c^2}\P \to \P$ is used for
convenience. By hypothesis this scales as $(\D^2)^{2\Delta_S - \gamma_{str}}$.
The critical regime for the number of monomers is achieved by tuning $K=K_c(1
-2\D^{2/\nu D} \s)$ where $1/\sqrt{K_c}$ is the critical (maximal) value of
$\P$
which
makes
the denominators vanish in (\ref{conv}) to lowest order in $\D$.
This singular behaviour of (\ref{conv})
allows
one to read off the scaling dimension.
It will now be computed for the various phases in turn.

\noindent {\bf High-Temperature Phase}
\newline From the analysis of the previous section
(\ref{lung})(\ref{ohlong}),
 at least for
the limits $c \to 1$ and $c \to (\sqrt{2}-1)^+$, the
scaling behaviour $p \sim \sqrt{\D}$ and $z \sim \D^2$ is appropriate
and each denominator in (\ref{conv}) is $O(p^2,\sqrt{z+\mu},\s)$.
By power counting one
has
\be
2\Delta_S - \gamma_{str} = {3S\over 4}
\eq
as occurs at infinite temperature \cite{kostov}.
{}From (\ref{new}) and $\nu D =2$ one
deduces $\gamma_{str}= -1/2$, which is known to be consistent with
a conformal field theory central charge $c=0$. Hence from
the dimensions in the plane (\ref{ddk}) $\Delta^{(0)} = (9L^2 -4)/96$
one finds the usual dilute
phase of self-avoiding random walks. It seems logical to assume that
this
behaviour is valid for the entire range $1 \geq c > \sqrt{2} -1$. Note
that
$\Delta_1$ is positive, meaning that the two ends of a single polymer
attract,
 while $\Delta_2$ is relevant but positive,
charateristic of a dilute regime. The
contact operators $\O_{2n}$, $n>1$, are all irrelevant.

\noindent {\bf Collapse Temperature}
\newline At $c= \sqrt{2} -1$  the scaling behaviour $p \sim \D$ and
$z \sim \D^2$ was appropriate and the denominators are
$O(p^2,z,\mu,\s)$. By power counting one finds
\be
2\Delta_S = \gamma_{str} + {S\over 2}
\eq
while $\nu D =1$ implies $\gamma_{str} = -1$ consistent with the
$c=-2$
conformal field theory at the end of the unitary minimal series
(though it
is not unitary). The scaling dimensions $\Delta^{(0)}= (L^2 -4)/16$ are
those of the usual dense phase of self-avoiding random walks in the plane for
which $R^2 \sim L$ \cite{dense}. $\O_1$ has negative dimension since the
dense polymer screens the endpoints causing them to repel, while
$\O_2$ is of dimension zero, like the identity operator
characteristic of a dense phase.  Of the contact
operators, $\O_4$ is relevant and $\O_6$ marginal.

\noindent {\bf Low-Temperature Phase}
\newline
Consider perturbing $c =\sqrt{2}-1 +\e$ for small $\e <0$ now.
Starting with $S=1$ one recalls from (\ref{comp}) that the $z$-integral
does not converge unless higher powers of
$u$
and $\mu$ in the expansion (\ref{long}) are taken into account, in which
case
there is some saddle point at $z \sim O(1/\D^2)$ corresponding to
values of the discrete variable $i \sim O(N)$. This is quite
odd behaviour but one can formally proceed perturbatively in $\e$.  The first
corrections have already been worked out in the exponent of (\ref{olong}).
Collecting all the significant terms at leading non-zero order in $\e$
\be
\psi \equiv a_1[\psi_0 -  \D^{2/\nu D} 9 \tilde{p}^2 -\a \e^3 \sqrt{z+\mu}\D
+(\mu -3(\mu + z)/2)\D^2 + \cdots] \label{max}
\eq
where the constant part $\psi_0 = 1 + O(\e^2)$. If $\e < 0$ however,
$a_1 \psi_0$ is now longer the maximal value of $\psi$ (which determines
$K_c$). To leading order in $\e$, the
maximal value now occurs not for $z \to 0$ but for $z = z_c
\sim O(\e^6/\D^2)$ (the term of order  $u^2$ is now comparable
to the $\e^3 u$ term).  The ellipses in
(\ref{max})  correspond to higher order terms in $\D$
and $\e$ taking this fact into account. In detail,
perturbing about this maximum as
$z = z_c + \e^3 \tilde{z}/\D$  one finds
$z_c = \a^2 \e^6 /9 \D^2$ and Gaussian fluctuations
\be
\psi \equiv a_1\left[\psi_0 + {\a^2 \e^6 \over 6} -
9 \D^2
\tilde{p}^{2}
+\mu \D^2 - {27 \tilde{z}^2 \D^2 \over 8 \a^2} +  \cdots
\right]
\eq
The scaling laws for $\tilde{p}$ and $\tilde{z}$ in the last
expression
have been fixed to pick up the lowest order in $\mu$, and $\nu D =1$
as a result. The  singular part of the $S=1$ correlator  is
then
\begin{eqnarray}
Z^0 < \int \O_1 \int \O_1 > & \sim & \e^3 \int_{-\infty}^{+\infty} d\tilde{p}
d\tilde{z} {1 \over
\s + 9 \tilde{p}^{2}
- \mu  +  {27 \tilde{z}^2  \over 8 \a^2}
} \\
&\sim & \e^3 \log{( \s - \mu)} \ \ , \ \ \s -  \mu \to 0
\end{eqnarray}
Similarly for $S>1$ one has singular multiple integrals
\begin{eqnarray}
Z^0 < \int \O_S \int \O_S > & \sim & (\e^3)^{S} \int_{-\infty}^{+\infty}
d\tilde{p}_1 \ldots d\tilde{p}_S d\tilde{z}_1 \ldots d\tilde{z}_S
{1 \over 2\s  - 2 \mu  +  9 \tilde{p}^{2}_{1}+ {27
\tilde{z}^{2}_{1}\over 8\a^2} + 9 \tilde{p}^{2}_{2}+ {27
\tilde{z}^{2}_{2}\over 8\a^2}} \cdots \nonumber \\
&& \cdots {1 \over 2\s  - 2 \mu  +  9 \tilde{p}^{2}_{S}+ {27
\tilde{z}^{2}_{S}\over 8\a^2} + 9 \tilde{p}^{2}_{1}+ {27
\tilde{z}^{2}_{1}\over 8\a^2}} \\
& \sim & \e^{3S} \log{(\s - \mu)} \ \ , \ \ \s - \mu \to 0
\end{eqnarray}

This implies $\Delta_S = 0$ for all $S$ and $\gamma_{str} =0$.
Logarithmic scaling violations are known to occur in the $c=1$
conformal
field theory of a free scalar field when
coupled to 2D quantum gravity precisely because of the fluctuating
metric tensor.
However
there is no obvious reason to expect this assignment of central
charge
in the present case; $\gamma_{str} =0$ does not determine it
uniquely. In fact the KPZ relation (\ref{ddk}) would imply
$\Delta_{S}^{(0)} =0$ independent of $c$. This means that all
operators
scale like the identity operator. In particular $\Delta_{1}^{(0)} =0$
means that the polymer endpoints move freely, the repulsion due to
dense polymer being compensated by the attraction due to the tendancy
of
the polymer to collapse. The result $\Delta^{(0)}_{2n}=0$, $n > 1$,
would seem to imply a dense regime for both monomers and contacts
between
monomers --- the
lattice
is filled many times by the polymer. To this leading order in $\e$ at least
the multiple covering of the lattice by the polymer is not enough to
change the $\nu$-exponent from its usual dense-phase value $R^2 \sim
L$ however. It seems plausible that the behaviour to first order in
$\e$
should extend a finite distance into the low-temperature phase but
this
remains a conjecture.

\section{Conclusions}

To summarize the results, using  random matrices to solve the
polymer folding model on a fluctuating two-dimensional
lattice and then translating
the
results to the regular lattice using KPZ scaling, a collapse
transition
was found from a dilute to dense regime. On the fluctuating lattice
the
transition was third order. Although there are no rigorous bounds,
all known examples of bulk phase transitions show up with
lower  order on the regular lattice; the collapse transition of a dilute
 polymer in two dimensions  is really like a boundary phase transition
but nevertheless one might guess that it is first or second order.
While the high-temperature phase
and collapse point correspond to the usual dilute and dense phases of
self-avoiding walks in the plane respectively, the understanding of
the
low temperature phase is still incomplete. Some arguments were
given
to suggest an ultra-compact regime in which the polymer covers the
lattice
many times  due to the possibility of multiple occupation of
links.
This latter feature is a particularly unphysical aspect of the  model
as
regards real polymers
and results in a different universality class of collapse transition
from  models where multiple
occupation of links is forbidden. The reformulation of such models in
terms
of percolating clusters \cite{dup} showed a $\Theta$-point with
behaviour intermediate between that of the standard dilute and dense
phases of self-avoiding walks.

The indirect technique of solving discrete statistical models on fluctuating
lattices
by using random matrices and quantum gravity in order to predict the
critical phenomena on regular lattices has received little attention before
in the literature. While a little circuitous, the techniques invloved are
relatively
straightforward, having been developed by high energy physicists in
other
contexts. It would be interesting to see how far one could push it in
the condensed matter context. There are various possible extensions to the
present work: it is easy to find (but not so easy to solve) random matrix
models which forbid multiple occupation of links; this paper set up
the formalism for a finite polymer fugacity $\Gamma$ and it would be
interesting to study the interplay of inter- and intra-polymer
interactions;
the polymer can also be given more exotic monomer content by using
more complicated matrix models. These extensions are presently under
investigation.

\vspace{10mm}
\noindent Acknowledgements: I have benefitted from valuable
discussions
with J.Cardy, I.Kostov, and especially S.Flesia. Part of the work was
done while visiting Service de Physique Th\'eorique de Saclay and the
Aspen Center for Physics.

\newpage
\begin{center}
APPENDIX
\end{center}
Consider the expectation
\be
 < (1-c^2)^L \Tr [ (\F + \a \P)^{2L}] >_{\G=0} \label{cop}
\eq
with respect to the potential $V(\F,\P)$ (\ref{pot}).
Expanding this operator in
$\a$
\begin{eqnarray}
\Tr [(\F + \a \P)^{2L}]& =& \Tr [\F^{2L} + \a(\F^{2L-1}\P +
\F^{2L-2}\P\F
+ \cdots) \nonumber \\
& & + \a^2(\F^{2L-2}\P^2 + \F^{2L-3}\P\F\P + \F^{2L-3}\P\P\F +\cdots)
\nonumber
\\
&&+ \cdots + \a^{2L} \P^{2L}]
\end{eqnarray}
one sees that this represents a co-polymer with random sequencing
of monomers of type $\P$
and
type $\F$, the latter being made out of the same material as the
background
lattice as it were, together with a fugacity $\a$ for the proportion
which are type $\P$. The weights of $V$ are such that monomers of the same
type
tend to attract one another while monomers of different types tend to
repel.
Unfortunately this is the opposite situation to that of the physically
interesting one when monomers are electrically charged for example,
but it can be
solved immediately by the change of variable $\a \P + \F = \P'$,
$\F = \F'$. After some rescalings the new potential $V(\F',\P')$ is of
the
same form as the old one with the change
\be
c \to {\a c + 1 \over \sqrt{\a^2 + 1 + 2 \a c}}
\eq
Therefore (\ref{cop}) can be evaluated using the results of the paper;
in
particular there is a collapse transition the temperature of which
reaches
zero when $\a^2 =-1 + 1/(\sqrt{2} -1)^2$, there being no collapse for
$\a$ smaller than this value.

\newpage
\begin{center}
FIGURE CAPTIONS
\end{center}
\noindent Figure 1. -- (i) Allowed configurations of the polymer on a square
lattice. The double occupation of a link at `a' and `b' for example incurs a
contact weight $1/c^2$. Point `d' is not counted as a contact. (ii)
The
polymer is not allowed to cross itself, as at `e'.

\noindent Figure 2. -- A portion of a two-dimensional random square
lattice (dotted lines). The solid lines construct the dual graph given
by the Feynman diagram expansion of the random matrix model.

\noindent Figure 3. -- (i) A portion of a two-dimensional random
square lattice with a hole of length $6$ dual to the vertex created
by $\P^{2L}$ in the particular case $L=3$. In the dual graph lines
associated with $\F$ are solid while lines associated with $\P$ are
chain lines.
(ii) Sewing up the hole in the random square lattice to form a seam
representing a polymer.
The seam could begin at `a', `b', or `c' to give distinct
polymer/lattice
configurations. Shown is the case corresponding to
sewing `b' to `f' and `c' to `e'.

\noindent Figure 4. -- The induced free energy per unit length $u\l$
estimated from eqs.(\ref{sum})(\ref{derive}) at $L=150$.
Broken lines are the leading perturbative results
around $c=1$
and
$c=\sqrt{2} -1$ for $L = \infty$.

\noindent Figure 5 . -- A watermelon network contributing to
the two-point function $<{\cal O}_S {\cal O}_S >$ for the particular
case $S=3$

\noindent Figure 6 . -- Constructing the configurations contributing
to $<{\cal O}_2 {\cal O}_2 >$. (i) Surfaces $S_1$ and $S_2$ have disc
topology
and the same boundary length (the figure is not to scale). (ii) Sewing the
boundaries together gives a distinct polymer/surface  configuration
for
each choice of points $P_1,P_2,P_3$.

\vfil
\end{document}